\def\eeq{\end{equation}}
\def\beq{\begin{equation}}
\def\bea{\begin{eqnarray}}
\def\eea{\end{eqnarray}}

\documentstyle[aps,epsfig]{revtex}

\begin{document}

\title{Generalization of the Kolmogorov-Sinai entropy: 
Logistic- and periodic-like dissipative maps at the chaos threshold}

\author{Ugur Tirnakli$^{1,2}$, Garin F.J. Ananos$^1$ and 
Constantino Tsallis$^{1,3,4}$}

\address{$^1$Centro Brasileiro de Pesquisas Fisicas, Rua Xavier Sigaud
150, 22290-180 Rio de Janeiro-RJ, Brazil\\
$^2$Department of Physics, Faculty of Science, Ege University, 35100 
Izmir, Turkey \\
$^3$Department of Physics, University of North Texas, P.O. Box 311427, 
Denton, Texas 76203, USA \\
$^4$Santa Fe Institute, 1399 Hyde Park Road, Santa Fe, New Mexico 87501, USA\\
tirnakli@sci.ege.edu.tr, fedorja@cbpf.br, tsallis@cbpf.br}

\maketitle

\begin{abstract}

We numerically calculate, at the edge of chaos, the time evolution of the 
nonextensive entropic form 
$S_q \equiv [1-\sum_{i=1}^W p_i^q]/[q-1]$ (with $S_1=-\sum_{i=1}^Wp_i \ln p_i$) 
for two families of one-dimensional dissipative maps, namely a logistic- and a 
periodic-like with arbitrary inflexion $z$ at their maximum. At $t=0$ we 
choose $N$ initial conditions inside one of the $W$ small windows in which 
the accessible phase space is partitioned; to neutralize large fluctuations 
we conveniently average over a large amount of initial windows.
We verify that one and only one value $q^*<1$ exists such that the 
$\lim_{t\rightarrow\infty} \lim_{W\rightarrow\infty}
\lim_{N\rightarrow\infty} S_q(t)/t$ is {\it finite}, 
{\it thus generalizing the (ensemble version of) Kolmogorov-Sinai 
entropy} (which corresponds to $q^*=1$ in the present formalism). 
This special, $z$-dependent, value $q^*$ numerically coincides, 
{\it for both families of maps and all $z$}, with the one previously found 
through two other independent procedures (sensitivity to the 
initial conditions and multifractal $f(\alpha)$ function).\\

\noindent
{\it PACS Number(s): 05.45.-a, 05.20.-y, 05.70.Ce}

\end{abstract}


\vspace{1.5cm}

\section{Introduction}

In the area of nonlinear dynamical systems, a number of studies have recently 
addressed the sensitivity to initial conditions, multifractality and the
behavior of the Kolmogorov-Sinai (KS) entropy of these systems. In this
context, it is worth mentioning that these attempts include both
dissipative (low-\cite{TPZ,costa,lyra,circle} and high-\cite{high}
dimensional maps, symbolic sequences\cite{symbolic}) and conservative
(long-ranged many-body Hamiltonians\cite{latora,celia}, conservative
maps\cite{latbar}) systems. Before describing the purpose of the present
work, we need to briefly review the sensitivity to the initial conditions
and the multifractality of the chaotic attractor.

Firstly, let us focus on the concept of the sensitivity to initial
conditions. As already well studied in the
literature\cite{TPZ,costa,lyra,circle}, for one-dimensional systems it is
convenient to introduce $\xi (t) \equiv \lim_{\Delta x(0)\rightarrow 0} 
\frac{\Delta x(t)}{\Delta x(0)}\;$, where $\Delta x(0)$ is the
discrepancy of the initial conditions at time $t=0$, and $\Delta x(t)$ its
time dependence. It can be shown that $\xi$ satisfies the differential
equation $d\xi/dt=\lambda_1\xi$, where $\lambda_1$ is the Lyapunov
exponent, thus $\xi(t)=\exp(\lambda_1 t)$. Consequently, if $\lambda_1<0$ 
($\lambda_1>0$) the system is said to be {\it strongly insensitive
(sensitive)} to the initial conditions (and intermediate rounding). On
the other hand, if $\lambda_1=0$, then the function $\xi$ is expected to
satisfy the diferential equation $d\xi/dt=\lambda_q \xi^q$, hence 
\beq
\xi(t)= \left[1+(1-q)\lambda_q t\right]^{1/(1-q)}\;\;\;\;\;\; 
(q\in {\cal R}),
\eeq
which recovers the standard exponential case for $q=1$,  whereas $q\neq1$
yields a power-law behaviour. If $q>1$ ($q<1$) and $\lambda_q<0$ 
($\lambda_q >0$) the system is said to be {\it weakly insensitive
(sensitive)} to the initial conditions. Although asymptotic power-law
sensitivity to the initial conditions has since long been
observed\cite{grass,politi,mori}, eq.(1)  (which in fact corresponds to
the power-law growth of the {\it upper-bound} of $\xi(t)$) provides in
principle a more complete description since $t$ does not need to satisfy
$t>>1$. At the onset of chaos (where it is $\lambda_1=0$), this upper
bound ($\xi\propto t^{1/1-q}$) allows us to estimate the value $q^*$ of
index $q$ for the map under consideration. This method has been  
successfully used for a variety of maps such as logistic\cite{TPZ}, 
$z$-logistic\cite{costa}, circle\cite{lyra} and $z$-circular\cite{circle}
maps. 

Secondly, we consider another interesting property of dynamical
systems: the geometrical aspects of the attractor at the
onset of chaos. In order to describe the scaling behavior of the
critical dynamical attractor it is convenient to introduce a
multifractal formalism\cite{multif1,multif2}. In this formalism, 
it is possible to introduce a partition function 
$\chi_Q(N)=\sum_{i=1}^N p_i^Q$, where $p_i$ represents the
probability on the $i$th box among the $N$ boxes of the measure 
(it is necessary to warn the reader that we use here $Q$ instead of the
standard notation $q$ of multifractal literature in order to avoid
confusion with the present index $q$). In the $N\rightarrow \infty$ 
limit, the contribution to this partition function is proportional to 
$N^{-\tau(Q)}$, which comes from a subset of all possible boxes whose
number scales as $N_Q\propto N^{f(Q)}$, where $f(Q)$ is the fractal
dimension of the subset. The content on each contributing box scales as
$P_Q\propto N^{-\alpha(Q)}$ and all these exponents are related by a
Legendre transformation $\tau(Q)=Q\alpha(Q)-f(Q)$. The multifractal
measure is then characterized by multifractal function $f(\alpha)$,
which reflects the fractal dimension of the subset with singularity
strength $\alpha$. At the end points of the $f(\alpha)$ curve, this
singularity strength is associated with the most concentrated [namely,
$\alpha_{min}=\lim_{Q \rightarrow \infty}\alpha(Q)$] and the most
rarefied [namely, $\alpha_{max}=\lim_{Q \rightarrow -\infty}\alpha(Q)$]
regions on the attractor. Recently, the scaling behaviour of these regions
has been used to estimate the power-law divergence of nearby trajectories
and a new scaling relation has been proposed\cite{lyra} as 
\beq
\frac{1}{1-q^*} = \frac{1}{\alpha_{min}} - \frac{1}{\alpha_{max}}
\;\;\; .
\eeq
This relation constitutes a completely different method for the
calculation of the index $q^*$. Previous 
works\cite{TPZ,costa,lyra,circle} have shown, for various map families,
that the results of these two abovementioned methods of calculating the
$q^*$ values are the same within a good precision.

We are now prepared to describe the purpose of the present work, i.e., a
specific generalization of the KS entropy $K_1$. For a chaotic dynamical
system, one can define this entropy as the increase, per unit time, of the
standard Boltzmann-Gibbs entropy  $S_1=-\sum_{i=1}^W p_i \ln p_i$.
Moreover, it is well-known that the KS entropy is deeply related to the
Lyapunov exponents since the Pesin equality\cite{pesin} states that, for
vast classes of nonlinear dynamical systems, $K_1=\lambda_1$ if 
$\lambda_1 > 0$ and $K_1=0$ otherwise. Strictly speaking, as we shall
detail later on, the KS entropy is defined in terms of a {\it single
trajectory} in phase space, using a symbolic representation of the
regions of a partitioned phase space. However, it appears that, in almost
all cases, this  definition can be equivalently replaced by one based on
an {\it ensemble} of initial conditions. This is the version we use
herein.

The marginal cases, i.e., those for which $\lambda_1=0$, include period
doubling and tangent bifurcations as well as the onset of chaos. For these
cases, a generalized version of the KS entropy $K_q$ has been
introduced\cite{TPZ} as the increase rate of a proper nonextensive entropy
\beq
S_q(t) = \frac{1- \sum_{i=1}^W \left[p_i(t)\right]^q}{q-1}\;\;\; .
\eeq
This nonextensive entropy enables a generalization of Boltzmann-Gibbs
statistical mechanics \cite{tsallis1}; it recovers the standard 
Boltzmann-Gibbs entropy $S_1= -\sum_{i=1}^W p_i \ln p_i$ in the
$q\rightarrow 1$ limit. A general review of the properties of this entropy
and related subjects can be found in\cite{BJP}. So, for the generalized
version of the KS entropy, it has been proposed\cite{TPZ} 
\beq
K_q \equiv \lim_{t\rightarrow\infty} \lim_{W\rightarrow\infty}
\lim_{N\rightarrow\infty} \frac{S_q(t)}{t}
\eeq
where $t$ is the time steps, $W$ is the number of regions in the
partition of the phase space and $N$ is the number of points that are
evolving with time. The Pesin equality itself is expected to be
generalizable as follows: $K_q=\lambda_q$ (or some appropriate average)
if $\lambda_q > 0$ and $K_q=0$ otherwise. 

These ideas have been used very recently by Latora et al\cite{latbar2} to
construct a {\it third} method for the calculation of the $q^*$ values.
They conjectured that (i) a special $q^*$ value exists such that $K_q$ is
finite for $q=q^*$, vanishes for $q>q^*$ and diverges for $q<q^*$, (ii)
this value of $q^*$ coincides with that coming from other two methods of
finding $q^*$ (namely, from eq.(1) and eq.(2)). Latora et al have checked
these conjectures with numerical calculations for the standard logistic
map and found that the growth of $S_q(t)$ is linear when the value of
index $q$ equals $q^*\simeq0.2445$, which strongly supports the point that
all three methods yield one and the same special value of the index
$q^*$! Although the results of Latora et al provide strong evidence in
favor of this scenario, it is no doubt necessary to study more general
maps along these lines in order to see whether the results of this method
reproduce those of the previous two methods. The aim of this work is to
address this question by studying the $z$-logistic and $z$-periodic maps.

\section{Numerical results}

We first recall the $z$-logistic map
\beq
x_{t+1} = 1 - a |x_t|^z \;\;\; ,
\eeq
where $1<z$, $0<a\le 2$, $-1\le x_t \le 1$, and the $z$-periodic map
\beq
x_{t+1} = d\; \cos\left(\pi\left|x_t-\frac{1}{2}\right|^{z/2}\right)
\;\;\; ,
\eeq
where $1<z$, $0<d<\infty$, $-d\le x_t \le d$. These maps display
a period-doubling route to chaos and the parameter $z$ is the inflexion
of the map at its extremal point. Both maps belong to the same
universality class, namely they share the  same value $q^*$  for a given
$z$. The values of $a_c$ and $d_c$ at the  onset of chaos, as well as the
$q^*$ values calculated from eq.(1) and eq.(2) are indicated in the Table
for three representative values of $z$.

Now we can describe the numerical procedure that we used as the third
method for calculating $q^*$. This method was first introduced
in\cite{latbar} for conservative systems and then used by Latora et
al\cite{latbar2} for the standard ($z=2$) logistic map. We partition the
interval $-1\le x\le 1$ ($-d\le x\le d$) into $W$ equal windows for the
$z$-logistic map ($z$-periodic map). Then we choose (randomly or not) one
of these windows and select (randomly or uniformly) $N$ initial values of
$x$ (all inside the chosen window) for the $z$-logistic ($z$-periodic) map
at a given value of $a$ ($d$). As $t$ evolves, these $N$ points typically
spread within the interval of phase space and this gives us a set
$\{N_i(t)\}$ with $\sum_{i=1}^W N_i(t)=N,\; \forall t$, which
consequently yields a set of probabilities $\{p_i(t)\equiv N_i(t)/N\}$.
Since for $t=0$ all $N$ points are inside the chosen window, $S_q(0)=0$
and, as time evolves, $S_q(t)$ starts increasing,  although bounded by the
equiprobability value (i.e., by $(W^{1-q}-1)/(1-q)$ for $q\neq 1$, and
$\ln W$ for $q=1$). Let us anticipate our main result. For $N$ and $W$
increasingly large (and always satisfying $N >>W$) we observe, in all
cases, that a {\it linear} increase of $S_q$ with time emerges only for a
special value of $q$, namely $q^*(z)$. We define the corresponding slope
as the generalized Kolmogorov-Sinai entropy $K_q$. For values of $a$ and
$d$ for which $\lambda_1>0$ we verify that $q^*=1$. But at the edge of
chaos, we obtain $q^*<1$.

In contrast with the cases where strong chaos exists,  considerably large
fluctuations appear in $S_q(t)$ at the chaos threshold, due to the fact
that the attractor occupies only a tiny part of the available phase
space. These fluctuations make uneasy the task of determining the value
of $q$ for which linearity is present. In order to neutralize these
fluctuations, we adopt a procedure of averaging over the "efficient" 
initial conditions as introduced in\cite{latbar2}. The procedure is as
follows: we choose the initial distribution of $N$ points in one of
the $W$ cells of the partition and count the total number of occupied
cells as time evolves between say $t=1$ and $t=50$; this number is a
measure of the efficiency of that particular window in spreading itself.
After studying each one of the $W/2$ cells in the interval $0\leq x\leq
1$ ($0\leq x\leq d$) for the $z$-logistic ($z$-periodic) maps, the
averaging is done over the cells for which the total number of
occupied cells is larger than a fixed threshold, say 5000 (see Fig. 1).

Our results for typical values of $z$ for the $z$-logistic and the
$z$-periodic maps at the threshold of chaos are indicated in Figs. 2-4.
We used $N=10 \times W=10^6$  for the $z$-logistic maps, and $N=10 \times
W=6\times 10^5$ for the $z$-periodic maps. The values we have used are
large enough so that the time evolution of the entropy in the intermediate
time region (after the initial transient and before the saturation) does
not depend much on $W$. This point has been illustrated in Fig.2, where
we plot the time evolution of $S_q(t)$ for two values of $W$ for the case
of the $z=1.75$ logistic map. It is clear from this figure that the
starting points of saturation of the curves shift to larger time  as $W$
increases, but in the intermediate time region the curves are almost
insensitive to $W$. 

In all cases, the growth of $S_q(t)$ in this intermediate time region 
is found to be linear when $q=q^*$, whereas it curves upwards for $q<q^*$
and downwards for $q>q^*$, i.e., $K_q$ is finite for $q=q^*$, diverges for
$q<q^*$ and vanishes for $q>q^*$. In order to provide quantitative support 
to this behavior, we fit the curves with the polynomial $S(t)=A+Bt+Ct^2$
in the interval $[t_1,t_2]$ characterizing the intermediate region. We 
define the nonlinearity coefficient $R\equiv C (t_1+t_2) / B$ as a
measure of the importance of the nonlinear term in the fit; $R$ vanishes
for a strictly linear fit (see the Insets of Figs. 2-4). The times $t_1$
and $t_2$ are respectively defined as the end of the initial transient
(during which $S_q$ is roughly constant) and the beginning of the
saturation; $R(q)$ is almost independent from $(t_1,t_2)$ for any set of
specific values $(t_1,t_2)$ which satisfy the above characterizations. We
also see in the insets that the signs of $R$ on both sides of $q^*$ are
consistent with the curvatures of $S_q(t)$ in the intermediate region.

\section{Conclusions}

Let us make a few general considerations in order to better expose the
meaning of the present effort and its results. The Kolmogorov-Sinai
entropy $K_1$ is an important concept in chaotic dynamical systems,
either dissipative (like the one-dimensional maps considered here) or
conservative (like classical many-body Hamiltonians, satisfying the
Liouville theorem). Its definition is based on a partition of the
accessible phase space in a set of $W$ subspaces that are visited, along
time, in some complex order starting from a single initial point in phase
space. If we associate to each subspace a symbol, we shall have $W^{{\cal
N}}$ possible words of length ${\cal N}$. These  $W^{{\cal N}}$ words are
visited along time with probabilities $\{\pi_l\}$ ($l=1,2,..., W^{{\cal
N}}$). This set of probabilities enable the calculation of the 
Boltzmann-Gibbs-Shannon entropy $S_1(\{\pi_l\})=-\sum_l \pi_l \ln \pi_l$. 
In the limit ${\cal N} \rightarrow \infty$,  $S_1(\{\pi_l\})$ is
proportional to ${\cal N}$ if the system mixes {\it exponentially}
quickly (i.e., positive Lyapunov exponents), and $K_1$ is defined as the
supremum of the $\lim_{{\cal N} \rightarrow \infty} S_1({\cal N})/{\cal
N}$. Although we are not aware of any general proof, a common belief
exists that this computationally quite heavy definition can be
conveniently replaced by the one we have used here, based on an ensemble
of initial conditions instead of on single trajectories. This is to say,
we choose $N$ initial conditions within one of the $W$ subspaces, and we
follow along time the set of probabilities $\{p_i\}$ ($i=1,2,...,W$)
associated with the occupancies of those subspaces. The set  $\{p_i\}$
enables the calculation of $S_1(\{p_i\}) = -\sum_i p_i \ln p_i$, and
$K_1$ is expected to be the supremum of the $\lim_{t \rightarrow \infty}
S_1(t)/t$ .

The scenario we have just described is the standard one. What we have
exhibited here is a step forward, although only for very simple cases,
namely the maps herein considered. By analyzing the generalization
\cite{TPZ} $K_q$ of the KS entropy based on the nonextensive entropy
$S_q$, either in its trajectory formulation or in its ensemble one, we
exhibit a path that might be applicable to situations much more general
than the maps focused on here. We have (numerically) shown that an unique
value $q^*$ exists such that $K_q$ is finite, being zero for $q>q^*$ and
infinite for $q<q^*$. If the nonlinear dynamical system is {\it strongly
chaotic} ({\it exponential} mixing in phase space), then $q^*=1$, thus
recovering the usual scheme. But if the system is only {\it weakly
chaotic} ({\it power-law} mixing, i.e., vanishing Lyapunov exponents, but
positive generalized Lyapunov exponents, like $\lambda_q$), then one  
essentially expects $q^*<1$. Let us emphasize that in the present work we
have only addressed $q^*$ and not $K_{q^*}$ (nor $\lambda_{q^*}$).   
Indeed, the averaging procedure we have implemented is expected to
preserve linearity with time, but certainly not the supremum, or whatever
analogous to that. Further analysis is certainly welcome.

As in the case discussed by Latora et al\cite{latbar2}, we have verified
that the value $q^*$ that has emerged is precisely the same previously
obtained through two completely different procedures, namely the
power-law sensitivity to the initial conditions (see eq. 1), and the
multifractal structure of the chaotic attractor (see eq. 2). This
uniqueness of the special value $q^*$ clearly provides to the whole
scenario a very robust consistency.

Let us end with a speculative question quite analogous to the one we have 
addressed here. If we consider two systems $A$ and $B$ that are
independent in the sense that $p_{ij}^{A+B}=p_i^Ap_j^B$, then we
straightforwardly verify that $S_q(A+B) = S_q(A) + S_q(B) + (1-q)
S_q(A)S_q(B)$. In other words, we have extensivity, superextensivity and
subextensivity for $q=1$, $q<1$ and $q>1$ respectively. Let us now assume
that we have a classical $d$-dimensional many-body Hamiltonian system
whose particles interact through two-body interactions which  are nowhere
singular and which (attractively) decay like $r^{-\alpha}$ at long 
distances. It is known that such a system is thermodynamically extensive
if $\alpha/d>1$ (short-range interaction) and nonextensive if $0 \le
\alpha/d \le 1$ (long-range interaction) \cite{BJP}. For $\alpha/d>1$, if
the system is thermodynamically large  (say it contains $M$ particles with
$M \simeq 10^{23}$), any two, also  thermodynamically large, of its
subsystems can be considered as probabilistically independent in the
above sense. Then, because of its well known extensivity, the special
value of $q$ to be associated clearly is $q=1$. Consequently the entropy
$S_1$ is generically expected to yield a {\it finite} value for $\lim_{t
\rightarrow \infty} \lim_{M \rightarrow \infty} S_1(t,M)/M$.  The question
we wish to leave open at the present stage is: for $0 \le \alpha/d \le
1$, is there a special (possibly unique) value of $q$ such that $\lim_{t
\rightarrow \infty} \lim_{M \rightarrow \infty} S_q(t,M)/M$ also is 
{\it finite}?

\section*{Acknowledgments}
One of us (UT) acknowledges the partial support of BAYG-C program 
by TUBITAK (Turkish Agency) and CNPq (Brazilian agency) as well as 
by the Ege University Research Fund under the project 
number 98FEN025. This work has also been partially supported by 
PRONEX/FINEP and FAPERJ (Brazilian Agencies).

\newpage

{\bf Figure and Table Captions}

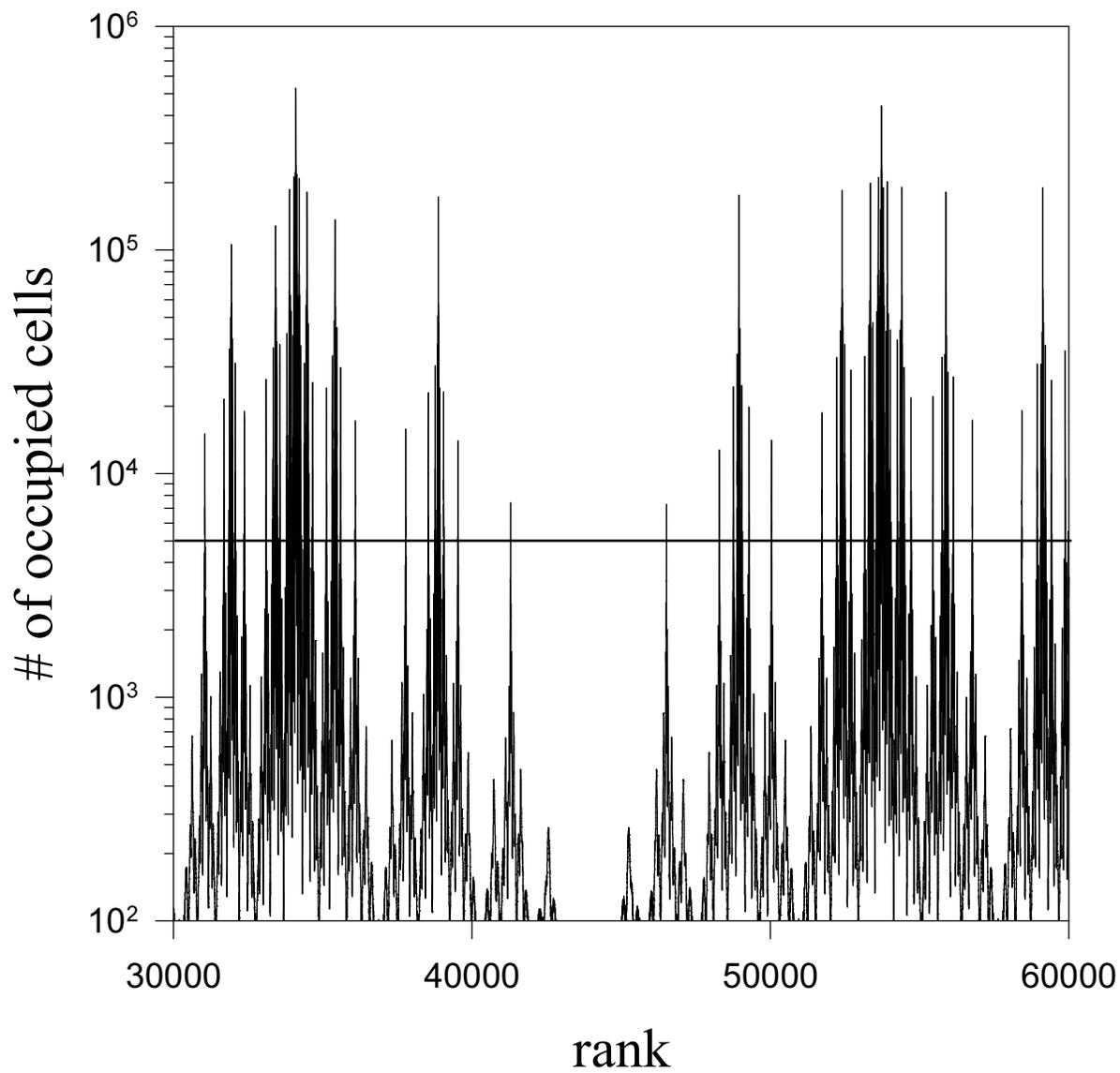
{\bf Figure 1} - Total number of occupied cells as function of the window
rank (varying from rank 30000 to rank 60000, respectively corresponding
to $x=0$ and $x=1$) for the $z=3$ periodic map with $W=6 \times 10^4$. 
The horizontal line indicates the cutoff at 5000. 
Average for $S_q(t)$ is done over the windows associated with values
above the cutoff.

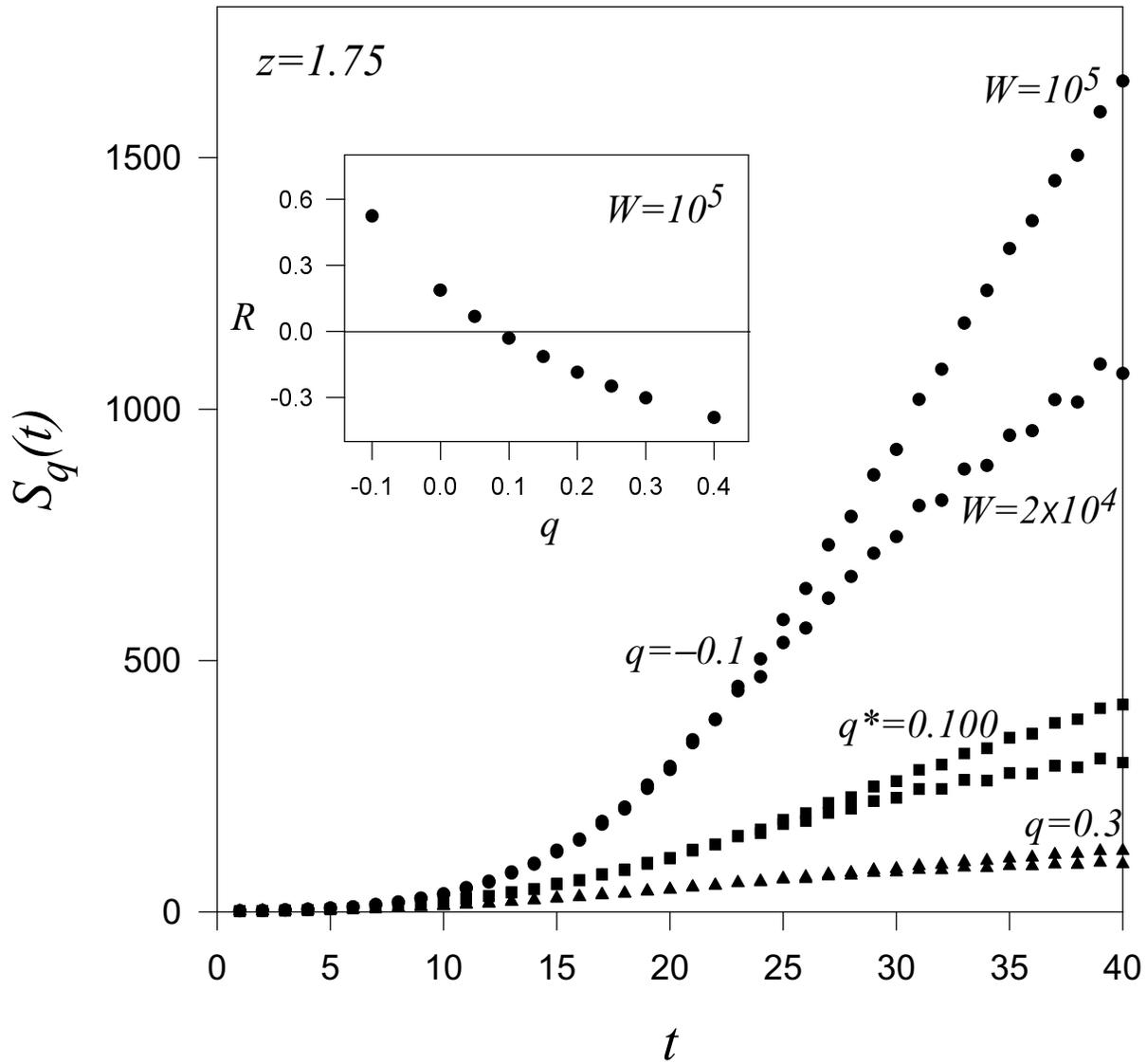
{\bf Figure 2} - Time evolution of $S_q(t)$ for two different values of
$W$ for the $z=1.75$ logistic map.
Inset: The nonlinearity coefficient $R$ versus $q$ (see text for
details).

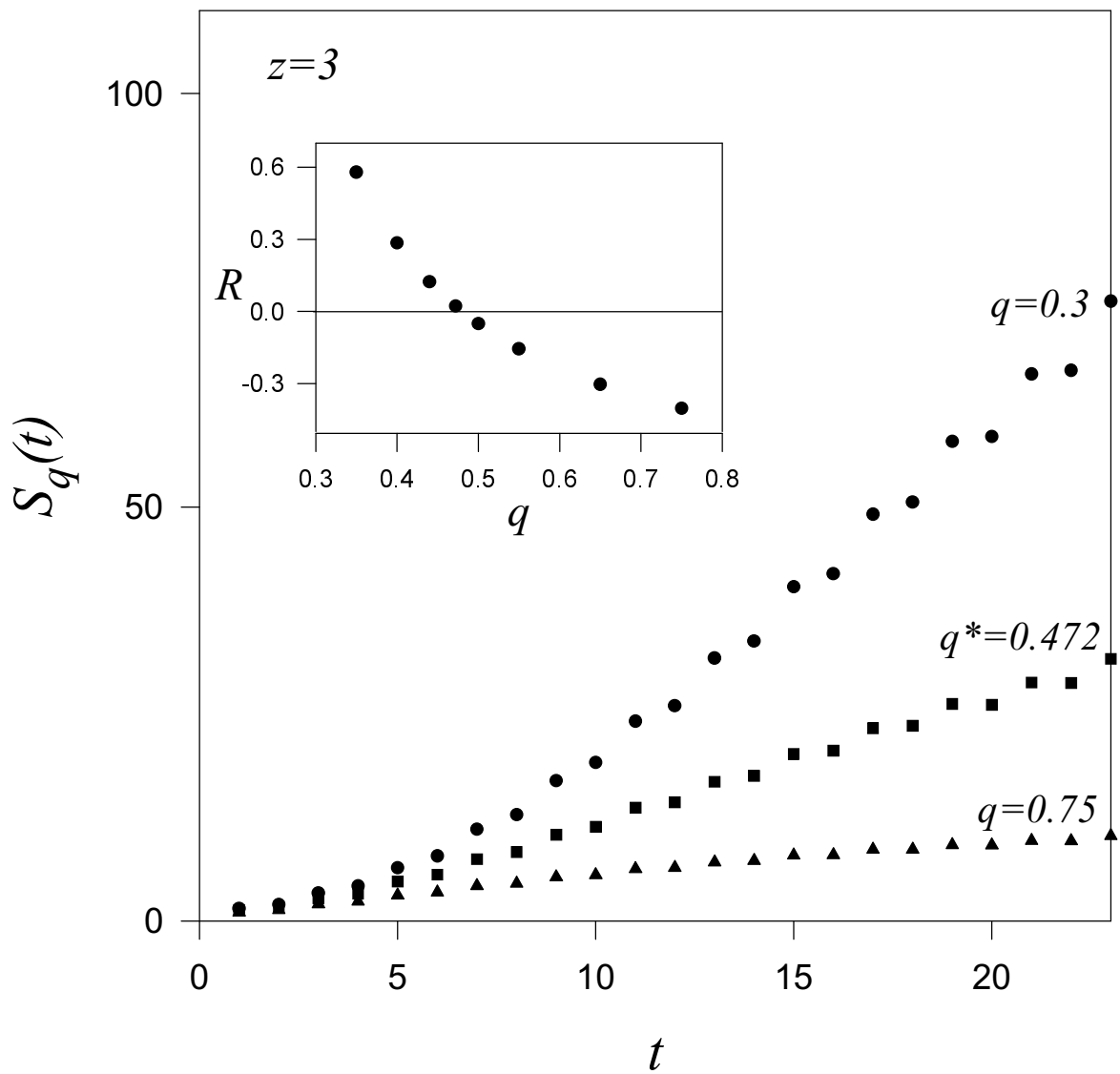
{\bf Figure 3} - Time evolution of $S_q(t)$ for three different values of
$q$ and $W=10^5$ for the $z=3$ logistic map. 
Inset: The nonlinearity coefficient $R$ versus $q$ (see text for
details).

{\bf Figure 4} - Time evolution of $S_q(t)$ for three different values of
$q$ and $W=6 \times 10^4$ for the $z$-periodic maps: (a) $z=1.75$; (b)
$z=2$; (c) $z=3$. Insets: The nonlinearity coefficient $R$ versus $q$ (see
text for details).

{\bf Table} - The values of $a_c$, $d_c$ and $q^*$ at the chaos threshold
for typical values of the inflexion parameter $z$.


\vspace{1.5cm}

\begin{center}
{\bf Table}
 
\vspace{1cm}

\begin{tabular}{||c|c|c|c||}  \hline
 & $z$-logistic map  & $z$-periodic map  &   \\
$z$  &       $a_c$       &    $d_c$          & $q^*$        \\
\hline
$\;1.75\;$  & $\;1.35506075...\;$ &$\;0.77946132...\;$ &$\;0.100\;$  \\ \hline
$\;2.0\;$   & $\;1.40115518...\;$ &$\;0.86557926...\;$ &$\;0.2445\;$ \\ \hline
$\;3.0\;$   & $\;1.52187879...\;$ &$\;1.07848805...\;$ &$\;0.472\;$   \\ \hline
\end{tabular}
 
\end{center}




\begin{thebibliography}{99}

\bibitem{TPZ} C. Tsallis, A.R. Plastino and W.-M. Zheng,
Chaos, Solitons and Fractals {\bf 8} (1997) 885.

\bibitem{costa} U.M.S. Costa, M.L. Lyra, A.R. Plastino and C. Tsallis,
Phys. Rev. E {\bf 56} (1997) 245.

\bibitem{lyra} M.L. Lyra and C. Tsallis, Phys. Rev. Lett. 
{\bf 80} (1998) 53.

\bibitem{circle} U. Tirnakli, C. Tsallis and M.L. Lyra, Eur. Phys. J. B 
{\bf 11} (1999) 309.

\bibitem{high} F.A. Tamarit, S.A. Cannas and C. Tsallis, Eur. Phys. J. B
{\bf 1} (1998) 545; A.R.R. Papa and C. Tsallis, Phys. Rev. E {\bf 57} 
(1998) 3923.

\bibitem{symbolic} M. Buiatti, P. Grigolini and L. Palatella, 
Physica A {\bf 268} (1999) 214.

\bibitem{latora} V. Latora, A. Rapisarda and S. Ruffo, Phys. Rev. Lett. 
{\bf 80} (1998) 692; {\bf 83} (1999) 2104.

\bibitem{celia} C. Anteneodo and C. Tsallis, Phys. Rev. Lett. {\bf 80} 
(1998) 5313.

\bibitem{latbar} V. Latora and M. Baranger, Phys. Rev. Lett. {\bf 82} 
(1999) 520.

\bibitem{grass} P. Grassberger and M. Scheunert, S. Stat. Phys. {\bf 26}
(1981) 697. 

\bibitem{politi} T. Schneider, A. Politi and D. Wurtz, Z. Phys. B {\bf
66}, 469 (1987); G. Anania and A. Politi, Europhys. Lett. {\bf 7} (1988)
119. 

\bibitem{mori} H. Hata, T. Horita and H. Mori, Progr. Theor. Phys. {\bf
82}, 897 (1989).

\bibitem{multif1} T.C. Halsey et al, Phys. Rev. A 33 (1986) 1141.

\bibitem{multif2} C. Beck and F. Schlogl, {\it Thermodynamics of chaotic
systems} (Cambridge University Press, Cambridge, 1993)

\bibitem{pesin} Ya Pesin, Russ. Math. Surveys {\bf 32} (1977) 55.

\bibitem{tsallis1} C. Tsallis, J. Stat. Phys. {\bf 52} (1988) 479; 
E.M.F. Curado and C. Tsallis, J. Phys. A {\bf 24} (1991) L69 
[corrigenda: {\bf 24}  (1991) 3187 and {\bf 25} (1992) 1019]; C. Tsallis,
R.S. Mendes and A.R. Plastino, Physica A {\bf 261}, 534 (1998).

\bibitem{BJP}A set of mini-reviews is available in {\it Nonextensive
Statistical Mechanics and Thermodynamics}, eds. S.R.A. Salinas and C.
Tsallis, Braz. J. Phys. {\bf 29} (1999)
[http://sbf.if.usp.br/WWW$_{-}$pages/Journals/BJP/Vol29/Num1/index.htm];
C. Tsallis, in {\it Nonextensive Statistical Mechanics and its
Applications}, eds. S. Abe and Y. Okamoto, {\it  Lecture Notes in
Physics}  (Springer-Verlag, Berlin, 2000), in press.

\bibitem{latbar2} V. Latora, M. Baranger, A. Rapisarda and C. Tsallis, 
{\it The rate of entropy increase at the edge of chaos}, 
(1999) preprint (cond-mat/9907412).

\end{thebibliography}
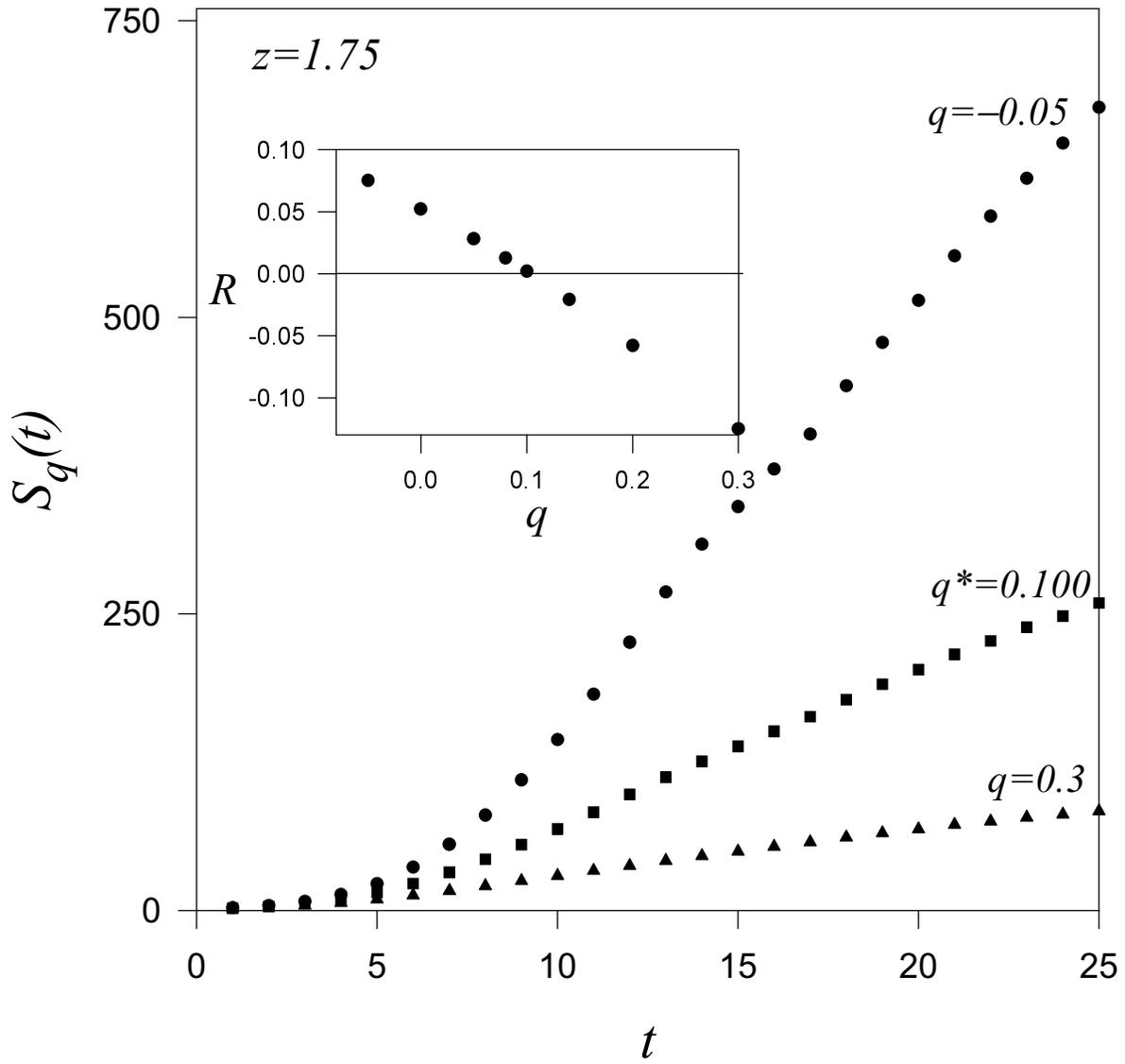
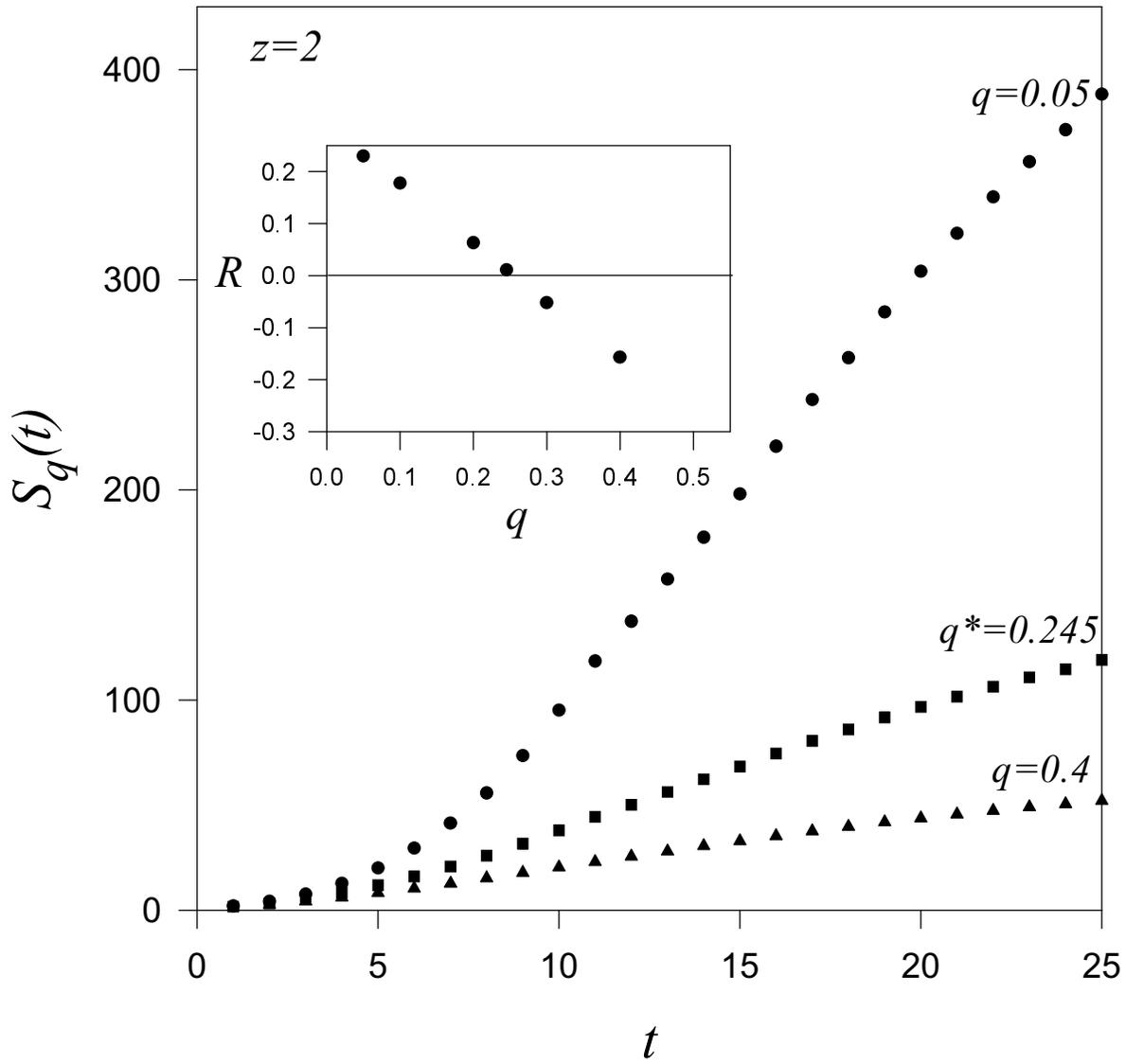
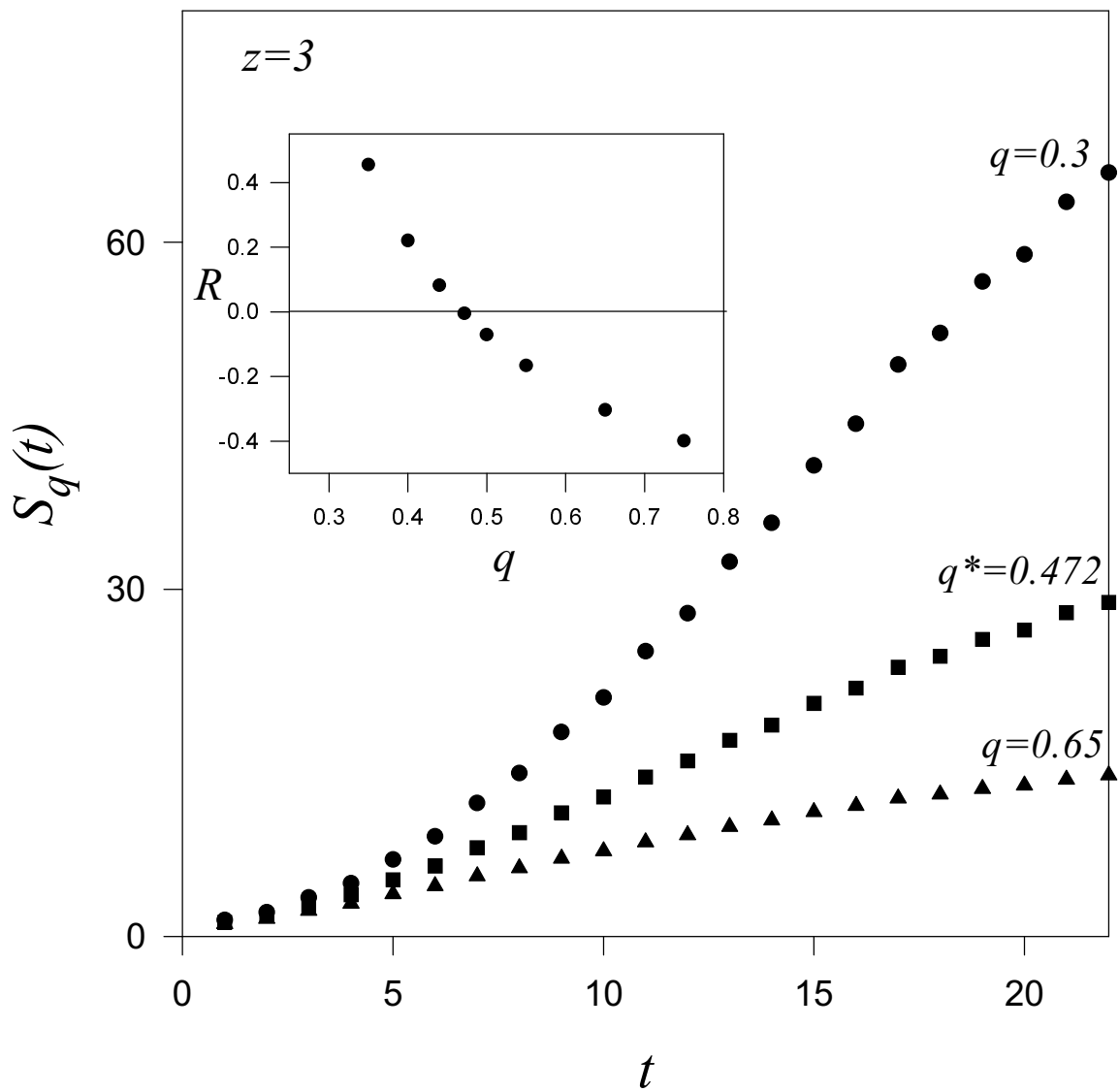
\end{document}